\documentclass[%
 reprint,
 amsmath,amssymb,
 aps,
]{revtex4-2}

\usepackage{graphicx}
\usepackage{dcolumn}
\usepackage{bm}
\usepackage{aas_macros}
\usepackage{amsmath}
\usepackage{comment}

\begin{document}

\title{Pair creation from radial electromagnetic perturbation of a compact astrophysical object}

\author{Mikalai Prakapenia$^{1,2}$}
\author{Gregory Vereshchagin$^{1,3,4,5}$}
\affiliation{$^{1}$ICRANet-Minsk, Institute of Physics, National Academy of Sciences of Belarus\\
220072 Nezalezhnasci Av. 68-2, Minsk, Belarus}
\affiliation{$^{2}$Department of Theoretical Physics and Astrophysics, Belarusian State University, Nezalezhnasci Av. 4, 220030 Minsk, Belarus}
\affiliation{$^{3}$ICRANet, 65122 Piazza della Repubblica, 10, Pescara, Italy}
\affiliation{$^{4}$ ICRA, Dipartimento di Fisica, Sapienza Universit\`a di Roma, Piazzale Aldo Moro 5, I-00185 Rome, Italy}
\affiliation{$^{5}$INAF -- Istituto di Astrofisica e Planetologia Spaziali, 00133 Via del Fosso del Cavaliere, 100, Rome, Italy}
\date{\today}

\begin{abstract}
Recently Usov's mechanism of pair creation on the surface of compact astrophysical objects has been revisited \cite{2024ApJ...963..149P} with a conclusion that the pair creation rate was previously underestimated in the literature by nearly two orders of magnitude. Here we consider an alternative hypothesis of pair creation due to a perturbation of the surface of a compact object. Radial perturbation is induced in hydrodynamic velocity resulting in a microscopic displacement of the negatively charged component with respect to the positively charged one. The result depends on the ratio between the spatial scale of the perturbation $\lambda$ and the mean free path $l$. When $\lambda\sim l$ the perturbation energy is converted into a burst of electron-positron pairs which are created in collisionless plasma oscillations at the surface; after energy excess is dissipated electrosphere returns to its electrostatic configuration. When instead $\lambda\gg l$, the perturbation is thermalized, its energy is transformed into heat, and pairs are created continuously by the heated electrosphere. We discuss the relevant astrophysical scenarios.
\end{abstract}

\maketitle

\section{Introduction}

Strong electric fields exceeding the Schwinger limit for pair production $E\sim E_c=m_{e}^2c^3/\hbar e \simeq 1.3\times 10^{16}$ V/cm, where $m_{e}$ is the electron mass, $e$ is its charge, $c$ is the speed of light and $\hbar$ is reduced Planck constant, may exist on bare surfaces of hypothetical quark stars \cite{1986ApJ...310..261A,1995PhRvD..51.1440K,1998PhRvL..80..230U,2005ApJ...620..915U,2006ApJ...643..318H,2010PhRvD..82j3010P,2024EPJC...84..463I,2024FrASS..1109463Z} or neutron stars \cite{2011PhLB..701..667R,2011PhRvC..83d5805R,2012NuPhA.883....1B,2014PhRvC..89c5804R}. The region with overcritical $E>E_c$ electric field in these objects is called \emph{electrosphere}. Such electrosphere at zero temperature is static and it does not produce electron-positron pairs due to the complete quantum degeneracy of electrons. The situation changes when the compact object is at a non-zero temperature.

Pair outflows from hot quark stars have been considered in \cite{1998PhRvL..80..230U,2001ApJ...550L.179U,2004ApJ...609..363A,2005ApJ...632..567A,2005ApJ...620..915U,2006ApJ...643..318H}. Recently \cite{2024ApJ...963..149P} by accurately treating dynamical pair creation it was shown that previous estimates of the luminosity in pairs can be underestimated by almost two orders of magnitude. It is indeed expected that newly formed compact objects may be hot and while most of the thermal energy is emitted via neutrinos \cite{2002PhRvL..89m1101P}, considerable energy may be emitted in pairs as well \cite{1998PhRvL..80..230U}.

An alternative to instant heating \cite{2001PhRvL..87b1101U} or continuous heating \cite{2001ApJ...559L.135U} may be a perturbation resulting in a relative displacement of negatively charged component (electrons) with respect to the positively charged component (protons or quarks). This is a form of radial perturbation of compact objects. Recently radial perturbations were considered e.g. in \cite{2019PhRvD.100k4041J,2020ApJ...897..168K,2021PhRvD.103j3003S,2023PhRvD.108f4007C}.

Electromagnetic perturbations were first considered in \cite{2010PhLB..691...99H,2012PhRvD..86h4004H} in the context of the gravitational collapse of baryon cores. Pair creation in such nonstationary electrosphere is expected as well. It was shown that the perturbation excites radial oscillations both in electron density, current and electric field. It was also indicated that electron-positron pairs are created in such nonstationary oscillating electrosphere.

In this work, we consider radial perturbations of electrosphere of compact astrophysical objects and study their relaxation for different parameters characterizing perturbations. With this goal relativistic kinetic equations with both source term describing Schwinger pair production and collisional term describing relaxation accounted for are solved numerically together with the Maxwell equations. In particular, we are interested in the spatial scale and amplitude of the perturbation. The paper is structured as follows. In Section \ref{sec2} electrostatic configuration is described. In Section \ref{dynamics} master equations are presented and initial conditions are described. In Section \ref{Pert} radial perturbation is introduced. In Section \ref{numres} we present our main results. Discussion and conclusions follow.

\section{Static configuration}\label{sec2}

First we describe the electrostatic configuration of fully degenerate electrons in a compact object possessing the electrosphere. As electrons are fully degenerate its distribution function $f_e$ represents a step function \footnote{Throughout the paper we adopt the system of units with $\hbar=c=k_B=1$.}.
\begin{equation}\label{fstep}
f_e= \left\{
\begin{array}[c]{cc}
1, & p < p_F,\\
0, & p > p_F,
\end{array}
\right.
\end{equation}
where $p_F$ is Fermi momentum. 

\begin{figure}[ht!]
\includegraphics[width=\columnwidth]{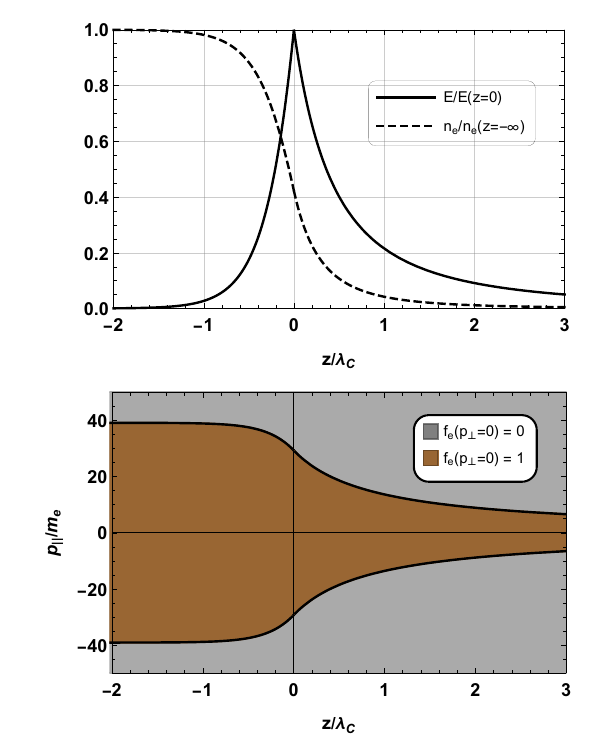}
\caption{Static solution of the Poisson equation \eqref{poisson} with the corresponding distribution function \eqref{fstep}.}
\label{figstaticnEf0}
\end{figure}
The electrochemical equilibrium condition for electrons is $\varepsilon_F=e\varphi$, where $\varepsilon_F$ is Fermi energy and $\varphi$ is electrostatic potential. The number density of electrons is \cite{1995PhRvD..51.1440K,2005ApJ...620..915U}
\begin{equation}
n_e=\int \frac{2}{(2\pi )^3} d^3p f_e =  \int\limits_0^{p_F} \frac{2}{(2\pi )^3}d^3p=\frac{p_F^3}{3\pi^2}.
\end{equation}

In ultrarelativistic approximation for electrons $p_F=\varepsilon_F$ and the Poisson equation gives \cite{1986ApJ...310..261A}
\begin{equation}\label{poisson}
    \frac{d^2\varphi}{dz^2}=-\frac{4\alpha}{3\pi}\left[e^2\varphi^3-n_q\right],
\end{equation}
where $n_q(z<0)=(\alpha/3\pi^2) (e^2\varphi_q^3)$ is the positive charge density, assumed to be described by a step function with the boundary at $z=0$, $z$ is spatial coordinate normal to electrosphere. The electric field $E(z)$ is defined from the electrostatic potential $\varphi(z)$ as $E(z)=-d\varphi/dz$.

For definiteness in what follows we assume that our compact object is a quark star. Quark number density function $n_q$ represents a step function and we use quark Fermi energy of 20 MeV: $\varepsilon_{q,F}=e\varphi_q = 20$ MeV \cite{1998PhRvL..80..230U}.  The electrostatic solution is presented in Figs. \ref{figstaticnEf0}. The sharp boundary of positively charged quarks (not shown) results in electron radial distribution that induces an overcritical electric field within a microscopic region near the boundary. As we are interested only in this region, in order to have a confined computational domain, we can safely neglect effects of spherical geometry and work in a plane geometry, so $z$ is considered as a Cartesian coordinate, which measures distance from the surface. Particle momentum can be decomposed into two parts: $p_{||}$ is parallel to $z$-axis and $p_\perp$ is orthogonal to $z$-axis.

Electron phase space distribution is also shown in Fig. \ref{figstaticnEf0}. As we consider particle dynamics in orthogonal direction to the surface of the compact object we introduce cylindrical coordinates in momentum space $\mathbf{p}=\{p_{\perp },p_\phi ,p_{||}\}$ with $p_{||}$-axis parallel to electric field $E$. Since $f_e=1$ for $p<p_F$ no phase space is available to create additional electrons in pairs with positrons, despite electric field is sufficiently large.

\section{Dynamical equations and collisional term}
\label{dynamics}

Particle evolution is described by one-particle electron/positron distribution function $f_\pm(t,z,p_{\perp },p_{||})$, which is normalized to the particle density $n_\pm=\int \frac{2 d^{3}p}{(2\pi )^{3}}f_\pm$.

We introduce dimensionless quantities $\tilde{t}=tm$, $\tilde{z}=zm$, $\tilde{p} =p/m$, $\tilde{E}=Ee/m_{e}^{2}$ to write the basic dynamic equations in dimensionless form. The set of Maxwell-Boltzmann-Vlasov equations in our case reduces to the Boltzmann-Amp\'ere system, which has to be supplemented with the Gauss law, as an initial condition. For this purpose, we use the Poisson equation \eqref{poisson}. Then we have
\begin{gather}
\label{vlasovampere}
\frac{\partial f_\pm}{\partial \tilde t} +\frac{\tilde p_{||}}{\tilde p^0}\frac{\partial f_\pm}{\partial \tilde z}\mp \tilde E\frac{\partial f_\pm}{\partial \tilde p_{||}} = \text{St}f_\pm \\ \notag
-(1-f_{+}-f_{-})|\tilde E|\text{ln}\left[1-\exp\left(  -\frac{\pi(1+\tilde p_\perp^2)}{|E|}\right)\right] \delta(\tilde p),   \\ \notag
\frac{\partial E}{\partial t} = 2 \alpha\int d^3 \tilde p \frac{\tilde p}{\tilde p^0}(f_- - f_+)+ \\ \notag
4 \alpha\frac{|\tilde E|}{\tilde E} (1-f_{+}-f_{-})\text{ln}\left[  1-\exp\left(-\frac{\pi(1+\tilde p_\perp^2)}{|\tilde E|}\right)  \right]\delta(\tilde p). 
\end{gather}
where St$f$ denotes the Boltzmann collision integral. The second term on the right hand side in both equations describes pair creation \cite{1987PhRvD..36..114G,2023PhRvD.108a3002P}.

Recently we considered pair creation by a hot electrosphere by solving the system of equations (\ref{vlasovampere}) and assuming $\text{St}f=0$, since electrosphere is essentially collisionless \cite{2024ApJ...963..149P}. However, below the surface boundary collisions are clearly important, as the density of electrons is high, see Fig. \ref{figstaticnEf0}.

To simplify the computational procedure instead of complete Boltzmann integral we apply the relaxation time approximation \cite{1954PhRv...94..511B,2017rkt..book.....V}, which reads 
\begin{equation}
\text{St}f=-\frac{f-f_0}{\tau}, 
\end{equation}
where $f_0$ is equilibrium distribution function and $\tau$ is characteristic thermalization timescale \footnote{Using the full collisional integral \cite{2007PhRvL..99l5003A} requires performing multidimensional integrals and is prohibitively computationally expensive}.

Since the Coulomb logarithm in relativistic plasma is of the order of unity \cite{2017rkt..book.....V}, the characteristic thermalization timescale $\tau$ for Coulomb interaction can be estimated using Thomson cross-section $\sigma_T$ as $\tau \simeq (\sigma_T n_e)^{-1}$ or in dimensionless form $\tilde\tau \simeq (\frac{8\pi}{3}\alpha^2 \tilde n_e)^{-1}$. In particular, within the star surface $\tilde n_e \approx 2000$ and $\tau\approx 1.1\times t_C.$ For the mean free path $l\simeq \tau$ we have then $l\simeq \lambda_C$.

\section{Radial electromagnetic perturbations}
\label{Pert}

Compact objects with electrosphere produce electron-positron outflows when  they are heated, as discussed by many authors \cite{1998PhRvL..80..230U,2003MNRAS.343L..69A,2024ApJ...963..149P}.

Alternatively, the energy may be deposited in the form of radial electromagnetic perturbation. Radial oscillations of quark stars have been studied recently in \cite{2019PhRvD.100k4041J,2020ApJ...897..168K,2021PhRvD.103j3003S,2023PhRvD.108f4007C}. Such perturbation can be induced e.g. by a collision with an external object \cite{2000ApJ...545L.127Z,2001PhRvL..87b1101U}, or by a decay of internal magnetic field, generating strong electric currents. While all previous works assumed that the perturbation is thermalized, here we study the nonequilibrium dynamics, which turns out to be rich: plasma oscillations decay and pair creation do operate on similar timescales.

We follow the idea introduced in \cite{2012PhRvD..86h4004H} and consider perturbation in bulk velocity of electrons $V_e$. This leads to the excitation of radial oscillations of the negatively charged component with respect to the positively charged one and the corresponding oscillations of electric fields and currents.

The radial bulk velocity $V_e$ is defined through the distribution function as
\begin{equation}
V_e = n_e^{-1} \int d^3p f_e p_{||} / p^0 ,  
\end{equation}
where $p^{0}=[p_{\perp }^{2}+p_{||}^{2}+m_{e}^{2}]^{1/2}$ is particle energy.

Perturbed distribution function of electrons can be written in the following form
\begin{gather}\label{initDF}
f_e= \left\{
\begin{array}[c]{cc}
1, & \left( ( p_{||} -\Delta )^2 + p_\perp^2 \right)^{1/2} < p_F,\\
0, & \left( ( p_{||} -\Delta )^2 + p_\perp^2 \right)^{1/2} > p_F,
\end{array}
\right.
\end{gather}
which contains some arbitrary function $\Delta(z)$ defined below. 

To study the interplay between energy generation and thermalization we consider two types of initial perturbations: 1) initial perturbation is localized within the mean free path $l$; 2) the spatial scale of initial perturbation is much larger than the mean free path $l$.

\begin{figure}[ht!]
\includegraphics[width=\columnwidth]{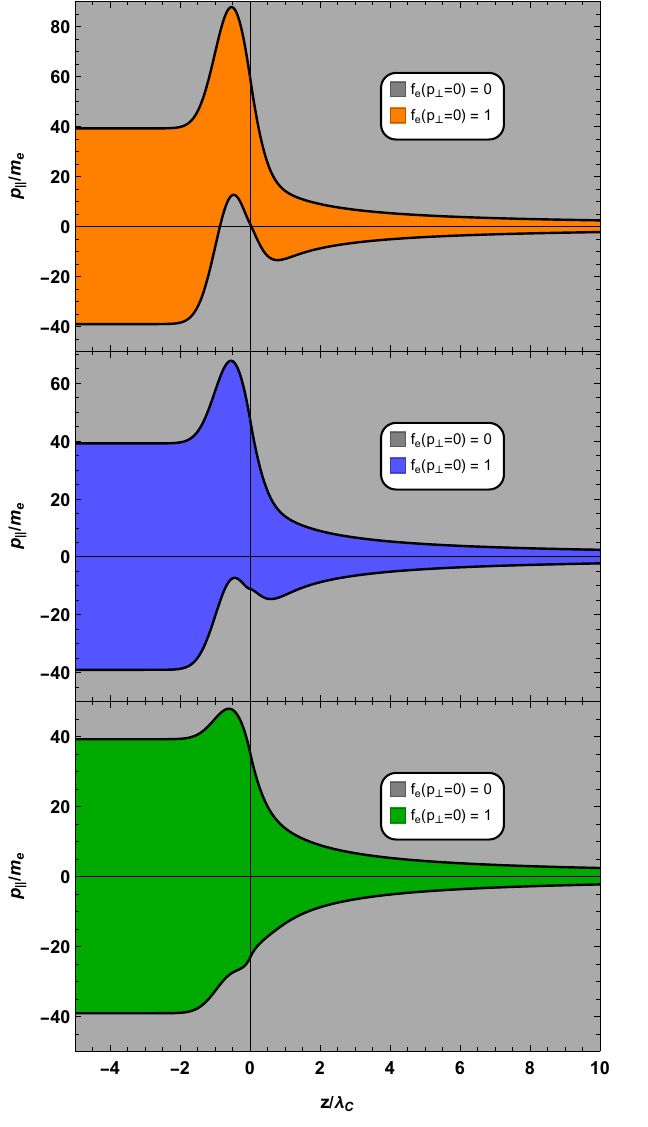}
\caption{Initial distribution function of electrons with $\Delta_1$. Orange: $\delta=50$. Blue: $\delta=30$. Green: $\delta=10.$}
\label{figinitDF}
\end{figure}

In the first case, $\lambda\approx\lambda_C\sim l$ and we adopt the following form of the perturbation
\begin{gather}
\Delta_1 =\delta \exp{(-2(z+0.5 \lambda_C)^2)}, 
\label{Delta1}
\end{gather}
where $\delta$ is the amplitude of initial perturbation measured in $m_e$ units. For $\delta=0$ the distribution function \eqref{fstep} is symmetric with zero bulk velocity $V_e=0$. Perturbation $\Delta_1$ is localized near the surface of the star where the particle density is sufficiently high. This perturbed distribution function of electrons is shown in Fig. \ref{figinitDF} for selected parameter $\delta$.

\begin{figure}[ht!]
\includegraphics[width=\columnwidth]{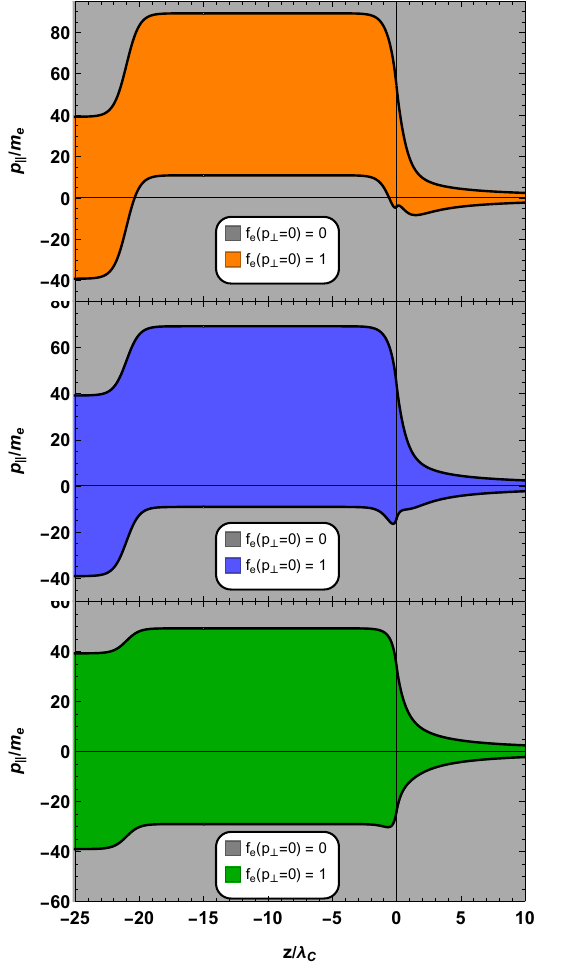}
\caption{Initial distribution function of electrons with $\Delta_2$. Orange: $\delta=50$. Blue: $\delta=30$. Green: $\delta=10.$}
\label{figinitDF2}
\end{figure}

In the second case, the perturbation is defined as
\begin{gather}
\Delta_2= \left\{
\begin{array}[c]{cc}
\frac{\delta}{2}[1+\tanh{(z+21\lambda_C)}], & z < -15 \lambda_C,\\
\frac{\delta}{2}[1+\tanh{(-z)}], & z > -15 \lambda_C,
\end{array}
\right.
\label{Delta2}
\end{gather}
which is again localized within the star, but on a large length scale $\lambda\simeq 20\lambda_C\gg l$. This distribution function is shown in Fig. \ref{figinitDF2} for selected parameter $\delta$.

Adding bulk velocity leads to an increase of energy density of electrons $\rho_e$, which is defined from the distribution function as
\begin{equation}
    \rho_e = \int \frac{2}{(2\pi )^3} d^3p f_e p^0.
\end{equation}
Maximal values of bulk velocity and energy density for both  $\Delta_1$ and $\Delta_2$ are shown in Table \eqref{tab1}.
\begin{table}[tbp]
\centering
\begin{tabular}[t]{|c|c|c|c|}
\hline
$\delta$          & $ 50 $  &  $30$    &  $10$  \\ \hline\hline
$\rho_{e,initial}/\rho_{e,static}$ & $2$ &  $1.5$ &   $1.1$  \\ \hline
$V_{e,initial}/c$ & $0.8$ &  $0.6$ &   $0.1$  \\ \hline
\end{tabular}%
\caption{Bulk velocity $V_e$ and energy density $\rho_e$ of perturbed electrons with initial profiles $\Delta_{1,2}$ for different values of $\delta$.}
\label{tab1}
\end{table}
These values imply that as the energy of pertirbation becomes comparable to the background energy, the bulk velocity becomes relativistic.

When initial perturbation is localized within the mean free path $l$ it will propagate in collisionless regime and can spread out. In this case one can set $f_0$ in form \eqref{fstep} corresponding to the static solution of the Poisson equation.

When the size of initial perturbation is much larger than the mean free path $l$ it will be affected by collisions. In this case we assume that the energy excess of perturbation $\rho_{e,initial}-\rho_{e,static}$ is transformed into thermal energy, so we set equilibrium distribution function $f_0$ in the form
\begin{equation}\label{FDdistr}
f_0= \frac{1}{1+\exp{\left[\left(\sqrt{{\tilde p}^{2} + 1}-{\tilde\mu}\right)/ \tilde\theta\right]}},
\end{equation}
where $\tilde\mu$ and $\tilde\theta$ are dimensionless chemical potential and temperature. In order to find $\tilde\mu$ and $\tilde\theta$ we solve the system of two equations: $n(\tilde\theta,\tilde\mu)=n_{e,initial}$ and $\rho(\tilde\theta,\tilde\mu)=\rho_{e,initial}$. For example, in the case of $\delta=50$ we have maximal temperature $\tilde\theta \simeq 15$ and the corresponding minimal chemical potential $\tilde\mu \ll \tilde \varepsilon_{q,F}$, which means that distribution function $f_0$ is highly non-degenerate.

\section{Numerical results}
\label{numres}

We integrate numerically the system of equations (\ref{vlasovampere}) and present the results for two different cases. 

\begin{figure}[ht!]
\includegraphics[width=\columnwidth]{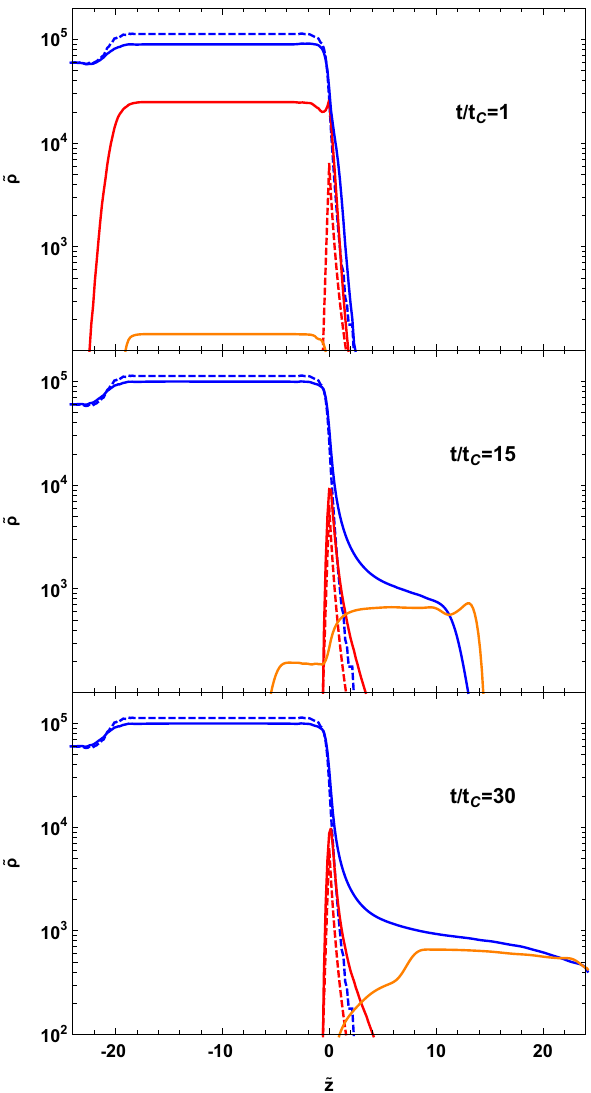}
\caption{Energy density for selected time moments.  The case of $\Delta_2$ with $\delta=50$. Blue: electrons. Orange: positrons. Red: electric field. Dashed lines show the initial values.}
\label{figrocol}
\end{figure}
\begin{figure}[ht!]
\includegraphics[width=\columnwidth]{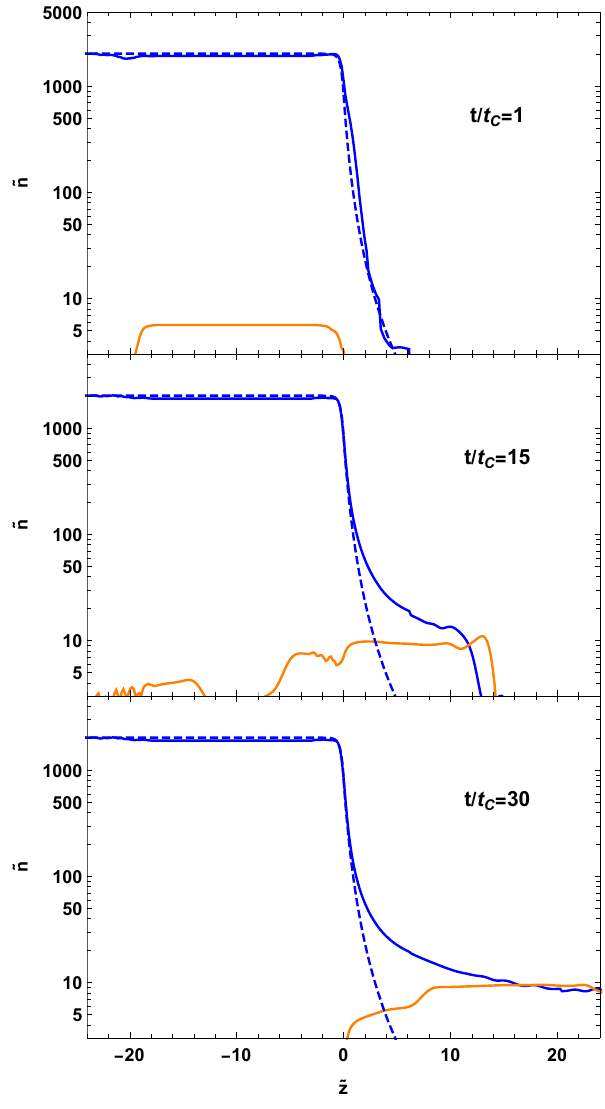}
\caption{Particle density for selected time moments. The case of $\Delta_2$ with $\delta=50$. Blue: electrons. Orange: positrons. Dashed lines show the initial values.}
\label{figncol}
\end{figure}
Fimensionless energy density for pairs is defined as
\begin{equation}
\tilde\rho_\pm = \int \frac{2}{(2\pi)^3}d^3\tilde p \tilde p^0 f_\pm,
\end{equation}
and for electric field as
\begin{equation}
\tilde\rho_E = \frac{\tilde E^2}{8\pi \alpha}.
\end{equation}

It is useful to integrate energy density over spatial coordinate $z$ to get total energy inside some space omitting a constant factor $4\pi R^2$ from the star surface area. We denote this dimensionless energy of type $i$ with a bar symbol
\begin{equation}
\bar{\rho_i} = \int d\tilde z \tilde \rho_i.
\end{equation}

First, we present the results for the initial perturbation (\ref{Delta2}) with the scale exceeding the mean free path. The corresponding initial distribution function is shown in the Fig. \ref{figinitDF2}. Below we discuss the case of $\delta=50.$ 

\begin{figure}
\includegraphics[width=\columnwidth]{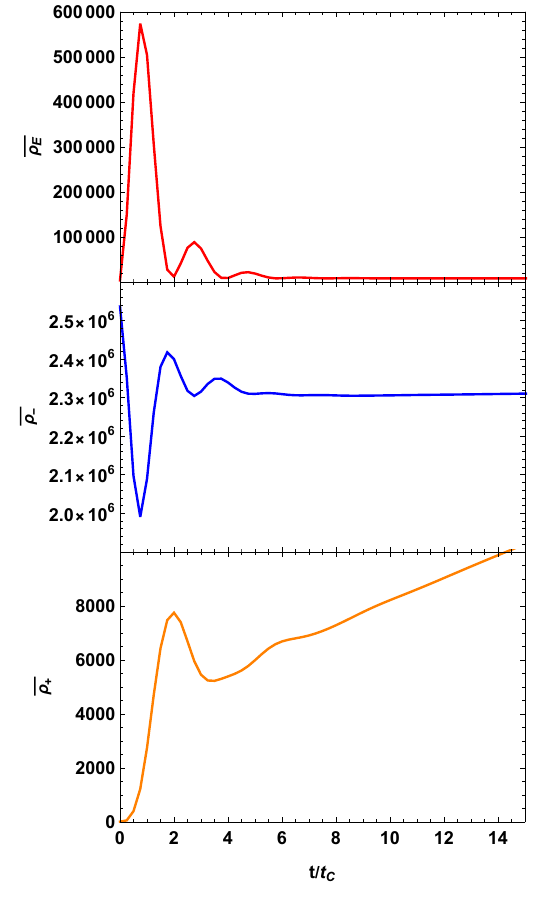}
\caption{Integrated energy density $\bar\rho$ for electric field and pairs as a function of time. The case of $\Delta_2$ with $\delta=50$. Blue: electrons. Orange: positrons. Red: electric field.}
\label{figE3col}
\end{figure}

Figs. \ref{figrocol} and \ref{figncol} show energy and particle densities at selected time moments. At the time moment $t=1 t_c$ electrons move outside the surface in accordance with its initial velocity, but energy excess is not dissolved yet. The perturbation in in fact a form of electric current which acts to increase the electric field within the surface, leading to electron-positron creation in that region. The integral energy in different components is shown in Fig. \ref{figE3col}. One can see that the increase of the electric energy occurs simultaneously with the decrease of the energy of electrons. Then at the time moment $t=15 t_c$ after a few oscillations, which are visible in Fig. \ref{figE3col}, electric field decreases down to its electrostatic values, and a flux of electron-positron pairs forms outside the surface of the star. At this point pairs are created in the electrosphere with nearly a constant rate, as can be seen also from Fig. \ref{figE3col}, where the energy of positrons grows monotonically. This pair creation process is caused by thermalization of perturbed electrons. Finally, at the time moment $t=30 t_c$ there is a steady approximately electrically neutral pair flux outside the surface.

\begin{figure}
\includegraphics[width=\columnwidth]{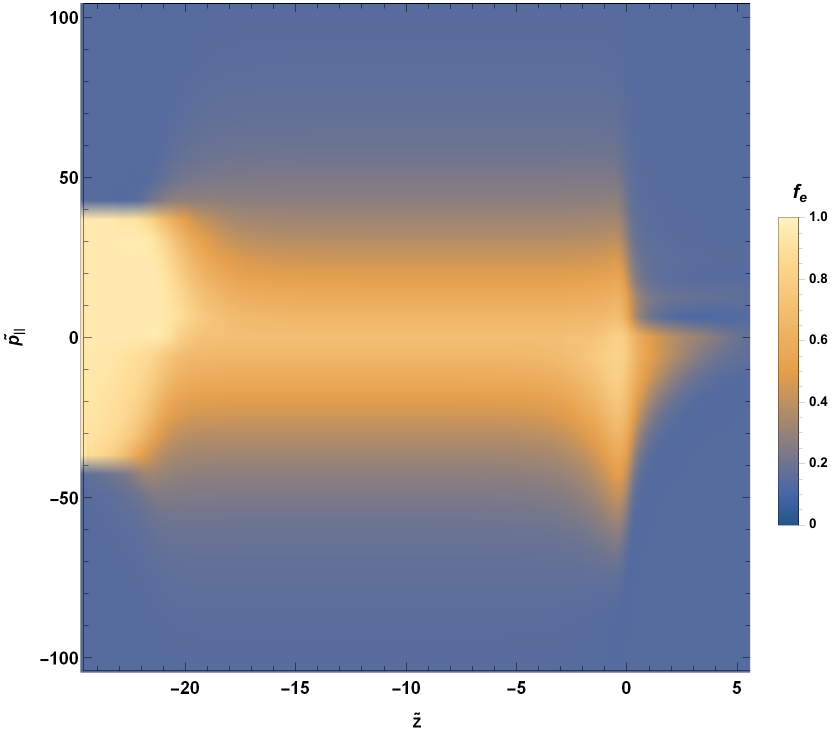}
\caption{Electron distribution function at time moment $t=30 t_C$. The case of $\Delta_2$ with $\delta=50$. }
\label{figfe28}
\end{figure}
The distribution function of electrons $f_e$ inside the star at this moment is shown in Fig. \ref{figfe28}, which clearly shows that initial energy excess of electrons (i.e. bulk velocity $V_e$) is transformed into thermal energy with a temperature $\tilde\theta\simeq 15$. Energy density of electrons inside the surface is close to its initial value, which indicates that collisions stopped the propagation of the perturbation. It is seen from Fig. \ref{figfe28} that distribution function of electrons inside the surface is symmetric in momentum and highly non-degenerate (i.e. $f_e<1$). Thus, the Pauli blocking in this spatial region is removed and pairs are produced in the same way as in the hot electrosphere considered in our previous work \cite{2024ApJ...963..149P}. Indeed, the energy flux and hence pair luminosity is consistent with those results. Such thermalization, i.e. conversion of kinetic energy of radial oscillations into thermal energy reminds the process observed first in \cite{2013PhLA..377..206B} in the homogeneous case.

Second, we present numerical results for the case when the spatial scale of perturbation is comparable to the mean free path $ \lambda\approx\lambda_C\sim l$. Here we do not expect thermalization to occur, hence we set the distribution function $f_0$ in collisional integral to be the same as in electrostatic solution.
The initial distribution function is shown in Fig. \ref{figinitDF}. Also in this case we discuss the solution with $\delta=50$.
\begin{figure}
\includegraphics[width=\columnwidth]{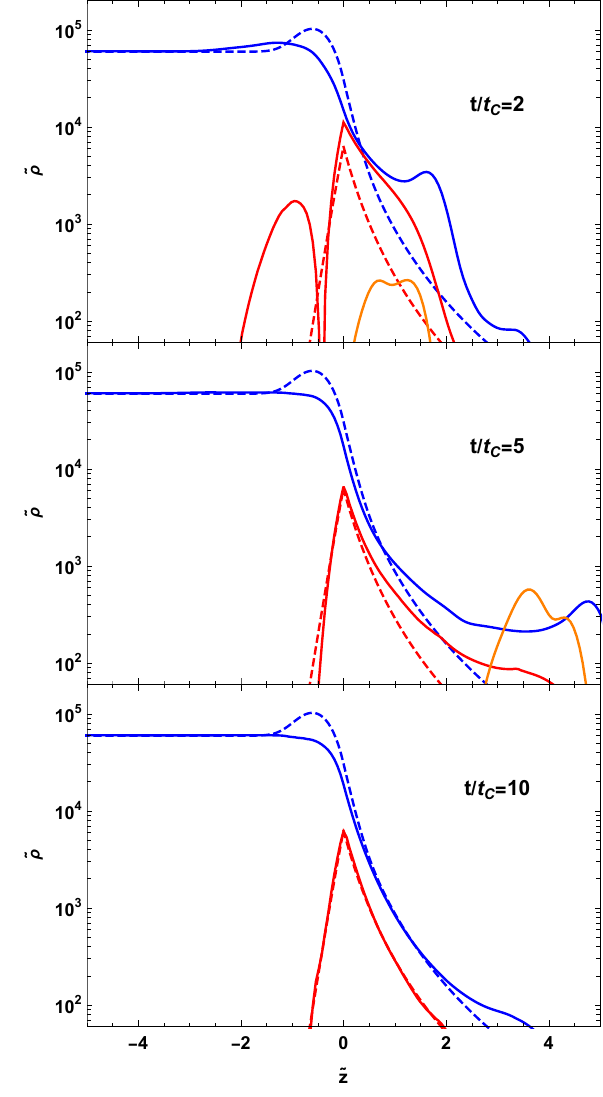}
\caption{Energy density for selected time moments.  The case of $\Delta_1$ with $\delta=50$. Blue: electrons. Orange: positrons. Red: electric field. Dashed lines show the initial values.}
\label{figrononcol}
\end{figure}
\begin{figure}
\includegraphics[width=\columnwidth]{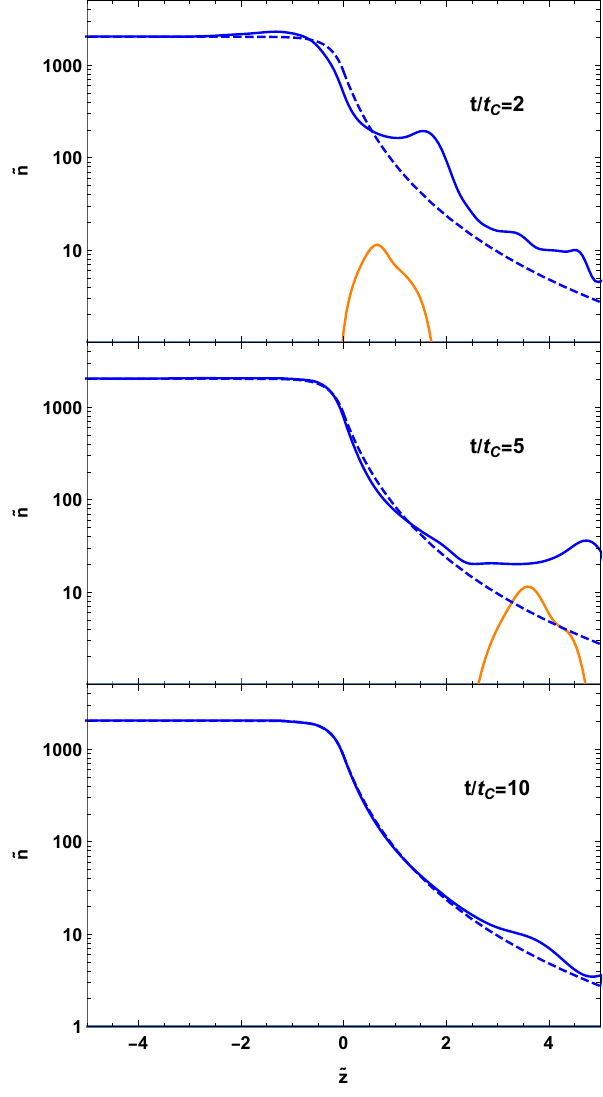}
\caption{Particle density for selected time moments. The case of $\Delta_1$ with $\delta=50$. Blue: electrons. Orange: positrons. Dashed lines show the initial values.}
\label{fignnoncol}
\end{figure}

Contrary to the previous case, initial energy excess of electrons disappears, as can be seen from Figs. \ref{figrononcol} and \ref{fignnoncol} already at the time moment $t=2t_C$. Due to electric current induced by the perturbation the electric field grows, as in the previous case. This causes a burst of pair creation. However, electric field then relaxes to its static configuration, and pair creation halts. At the time moment $t=5t_C$ there is a bunch of pairs which goes away and at  $t=10t_C$ we observe static configuration without additional pairs. Electron distribution function is also fully degenerate corresponding to the static configuration, as shown in Fig. 10. Thus, narrow energy excess with $\lambda\simeq l$ of the initial perturbation can not support Schwinger process continuously, but it works as a "gun" shooting a pair "bullet" with a size of a few Compton wavelength. Interestingly, pair creation in this case occurs on the background of almost degenerate electrons, but it is still possible due to plasma oscillations, as found in \cite{2023PhRvD.108a3002P}.
\begin{figure}
\includegraphics[width=\columnwidth]{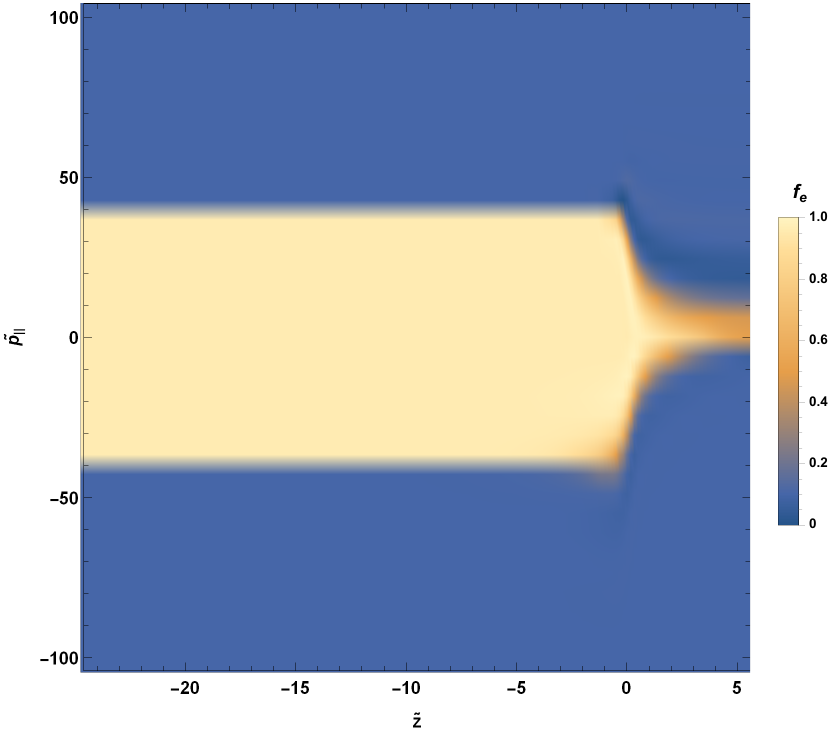}
\label{figstaticnEf01}
\caption{Electron distribution function at time moment $t=10 t_C$. The case of $\Delta_1$ with $\delta=50$. }
\end{figure}

Our results show that radial plasma oscillations are strongly dumped due to particle collisions. To see this we run simulations without collisional integral. In that case we obtained much longer oscillations, propagating mainly into the interior of the star, similar to what was found previously within the hydrodynamic approximation \cite{2012PhRvD..86h4004H}. Note however, that ideal hydrodynamics cannot take into account the dissipative effect of particle collisions, which is described in our simulation by the collisional integral.

For the second case, we also performed simulations assuming the distribution function $f_0$ is given by (\ref{FDdistr}) instead of \eqref{fstep}. In that case the energy remains in the perturbed region and pairs are again produced with a constant rate. That is expected, because with this assumption in essence the perturbation is no longer impulsive, but becomes stationary. Energy is being pumped into electrosphere continously, the duration of the process is sufficient for thermalization, despite the region when this actually happens is smaller than the mean free path.

Finally, we performed simulations varying the amplitude of perturbation $\delta$. The results appear to be qualitatively the same. Larger amplitude of perturbation produces larger currents and stronger electric field, which determines the amount of produced pairs. It is worth noting that due to specific form of the distribution function of perturbed electrons some of them eventually leave the system, overcoming electrostatic attraction, and some return to the surface and participate in the thermalization. The larger the amplitude of perturbation, the larger is the amount of electrons leaving the system.

\section{Discussion and conclusions}
\label{diss}

Impulsive or continuous heating were proposed \cite{2001PhRvL..87b1101U,2001ApJ...559L.135U} as a mechanism which launches pair creation by otherwise static electrosphere of a compact object. However, the detailed mechanism how this happens was not discussed. Qualitative arguments used in \cite{1998PhRvL..80..230U} underestimated the effect by nearly two orders of magnitude \cite{2024ApJ...963..149P}, while diffusion approximation adopted in \cite{2001ApJ...559L.135U} does not take nonequilibrium effects into account properly. Similarly, hydrodynamic approach adopted in \cite{2010PhLB..691...99H,2012PhRvD..86h4004H} does not take into account energy dissipation due to thermalization \cite{2007PhRvL..99l5003A,2009PhRvD..79d3008A} and grossly overestimates the duration of plasma oscillations.

In this work, we adopt the first principles approach and perform relativistic kinetic simulations describing spacetime evolution of the distribution function of electrons, positrons, as well as electric field self-consistently. This approach allows for the first time fully taking into account such effects as: Pauli blocking, particle collisions, pair creation by the Schwinger process, as well as backreaction of particles on electric field in electrosphere.
Our results show that all these phenomena indeed take place on a similar timescale.

As expected, perturbation of the electrostatic solution results in pair creation. Our simulations imply that the necessary condition for steady pair creation to occur is that the spatial scale of perturbation exceeds the mean free path, so that the electromagnetic perturbation is not dissipated, but its energy is converted into thermal energy, which in turn leads to opening up of the phase space for electrons and the corresponding possibility for prolific pair creation due to the Schwinger process. When this condition is not satisfied, pairs are created only in a short burst due to plasma oscillations, which quickly dissipate due to collisions. The mechanism of pair creation in this case is different from the thermal one: it is due to plasma oscillations, and not due to heating.

Our results are of interest in various astrophysical or cosmological settings. Quark droplets are considered as a viable candidate for dark matter \cite{diclemente2024strangequarkmatterdark} and their interactions in cores of galaxies may lead to pair production, and hence appearance of annihilation line, such as the one observed in the Galactic Center \cite{2005MNRAS.357.1377C}. Recently, a model of quark star crust collapse was proposed to explain the repeating fast radio bursts \cite{2021Innov...200152G}. In this scenario, an assumption of conversion of crust gravitational energy into heat was adopted. Our results clarify the microphysical evolution of the electrosphere perturbed by the crust collapse.

Such an electromagnetic perturbation may also originate either by infall of an external object such as a comet or a planet \cite{2001PhRvL..87b1101U} or by a decay of magnetic field, which results in induced electric currents. Considering that the thermal emission from quark matter is suppressed \cite{1991NuPhS..24...40C,2003ApJ...596..451C}, see, however \cite{2010PhLB..690..250Z}, pair creation appears to be the unique mechanism that drives dissipation of the energy of the perturbation.

 In the present paper, we focus on the spherically symmetric quark star and study the microphysical properties of its electrosphere. In the case where a magnetic field $\bf B$ is also present within the electrosphere, it enhances pair production according to the Schwinger formula $\dot n\propto \varepsilon\beta\exp(-\pi E_c/\varepsilon)/\tanh(\pi\beta/\varepsilon)$, where $\beta$ and $\varepsilon$ are electromagnetic field invariants $\varepsilon\beta={\bf E\cdot B}$ and $\varepsilon^2-\beta^2={\bf E^2-B^2}$, see \cite{2010PhR...487....1R}. The presence of macroscopic magnetic field outside electrosphere may lead to electromagnetic cascades \cite{1982ApJ...252..337D}. At large luminosities and large distances thermalization due to collisions occurs \cite{2004ApJ...609..363A}. We are at present considering the role of magnetic field on plasma evolution, but this topic is outside the scope of the paper.

{\bf Acknowledgements.} This work is supported within the joint BRFFR-ICRANet-2023 funding programme under the grant No. F23IKR-001.

\bibliography{total}{}

\hyphenation{Post-Script Sprin-ger}
\begin{thebibliography}{10}

\bibitem{2024ApJ...963..149P}
Mikalai {Prakapenia} and Gregory {Vereshchagin}.
\newblock {Pair Creation in Hot Electrosphere of Compact Astrophysical
  Objects}.
\newblock {\em \apj}, 963(2):149, March 2024.

\bibitem{1986ApJ...310..261A}
C.~{Alcock}, E.~{Farhi}, and A.~{Olinto}.
\newblock {Strange stars}.
\newblock {\em \apj}, 310:261--272, November 1986.

\bibitem{1995PhRvD..51.1440K}
Ch. {Kettner}, F.~{Weber}, M.~K. {Weigel}, and N.~K. {Glendenning}.
\newblock {Structure and stability of strange and charm stars at finite
  temperatures}.
\newblock {\em \prd}, 51(4):1440--1457, February 1995.

\bibitem{1998PhRvL..80..230U}
V.~V. {Usov}.
\newblock {Bare Quark Matter Surfaces of Strange Stars and $e^{+}e^{-}$
  Emission}.
\newblock {\em \prl}, 80:230--233, January 1998.

\bibitem{2005ApJ...620..915U}
V.~V. {Usov}, T.~{Harko}, and K.~S. {Cheng}.
\newblock {Structure of the Electrospheres of Bare Strange Stars}.
\newblock {\em \apj}, 620:915--921, February 2005.

\bibitem{2006ApJ...643..318H}
T.~{Harko} and K.~S. {Cheng}.
\newblock {Electron-Positron Pair Production in the Electrosphere of Quark
  Stars}.
\newblock {\em \apj}, 643(1):318--331, May 2006.

\bibitem{2010PhRvD..82j3010P}
Rodrigo {Pican{\c{c}}o Negreiros}, Igor~N. {Mishustin}, Stefan {Schramm}, and
  Fridolin {Weber}.
\newblock {Properties of bare strange stars associated with surface electric
  fields}.
\newblock {\em \prd}, 82(10):103010, November 2010.

\bibitem{2024EPJC...84..463I}
Adamu {Issifu}, Franciele~M. {da Silva}, and D{\'e}bora~P. {Menezes}.
\newblock {Proto-strange quark stars from density-dependent quark mass model}.
\newblock {\em European Physical Journal C}, 84(5):463, May 2024.

\bibitem{2024FrASS..1109463Z}
Xiao-Li {Zhang}, Yong-Feng {Huang}, and Ze-Cheng {Zou}.
\newblock {Recent progresses in strange quark stars}.
\newblock {\em Frontiers in Astronomy and Space Sciences}, 11:1409463, August
  2024.

\bibitem{2011PhLB..701..667R}
M.~{Rotondo}, Jorge~A. {Rueda}, R.~{Ruffini}, and S.~S. {Xue}.
\newblock {The self-consistent general relativistic solution for a system of
  degenerate neutrons, protons and electrons in
  {\ensuremath{\beta}}-equilibrium}.
\newblock {\em Physics Letters B}, 701(5):667--671, July 2011.

\bibitem{2011PhRvC..83d5805R}
M.~{Rotondo}, Jorge~A. {Rueda}, Remo {Ruffini}, and S.~S. {Xue}.
\newblock {Relativistic Thomas-Fermi treatment of compressed atoms and
  compressed nuclear matter cores of stellar dimensions}.
\newblock {\em \prc}, 83(4):045805, April 2011.

\bibitem{2012NuPhA.883....1B}
R.~{Belvedere}, D.~{Pugliese}, J.~A. {Rueda}, R.~{Ruffini}, and S.-S. {Xue}.
\newblock {Neutron star equilibrium configurations within a fully relativistic
  theory with strong, weak, electromagnetic, and gravitational interactions}.
\newblock {\em Nuclear Physics A}, 883:1--24, June 2012.

\bibitem{2014PhRvC..89c5804R}
Jorge~A. {Rueda}, Remo {Ruffini}, Yuan-Bin {Wu}, and She-Sheng {Xue}.
\newblock {Surface tension of the core-crust interface of neutron stars with
  global charge neutrality}.
\newblock {\em \prc}, 89(3):035804, March 2014.

\bibitem{2001ApJ...550L.179U}
V.~V. {Usov}.
\newblock {Thermal Emission from Bare Quark Matter Surfaces of Hot Strange
  Stars}.
\newblock {\em \apjl}, 550:L179--L182, April 2001.

\bibitem{2004ApJ...609..363A}
A.~G. {Aksenov}, M.~{Milgrom}, and V.~V. {Usov}.
\newblock {Structure of Pair Winds from Compact Objects with Application to
  Emission from Hot Bare Strange Stars}.
\newblock {\em \apj}, 609:363--377, July 2004.

\bibitem{2005ApJ...632..567A}
A.~G. {Aksenov}, M.~{Milgrom}, and V.~V. {Usov}.
\newblock {Pair Winds in Schwarzschild Spacetime with Application to Hot Bare
  Strange Stars}.
\newblock {\em \apj}, 632:567--575, October 2005.

\bibitem{2002PhRvL..89m1101P}
D.~{Page} and V.~V. {Usov}.
\newblock {Thermal Evolution and Light Curves of Young Bare Strange Stars}.
\newblock {\em Physical Review Letters}, 89(13):131101, September 2002.

\bibitem{2001PhRvL..87b1101U}
V.~V. {Usov}.
\newblock {Strange Star Heating Events as a Model for Giant Flares of
  Soft-Gamma-Ray Repeaters}.
\newblock {\em Physical Review Letters}, 87(2):021101, July 2001.

\bibitem{2001ApJ...559L.135U}
V.~V. {Usov}.
\newblock {The Response of Bare Strange Stars to the Energy Input onto Their
  Surfaces}.
\newblock {\em \apjl}, 559:L135--L138, October 2001.

\bibitem{2019PhRvD.100k4041J}
Jos{\'e}~C. {Jim{\'e}nez} and Eduardo~S. {Fraga}.
\newblock {Radial oscillations of quark stars from perturbative QCD}.
\newblock {\em \prd}, 100(11):114041, December 2019.

\bibitem{2020ApJ...897..168K}
Marek {Kutschera}, Joanna {Ja{\l}ocha}, {\L}ukasz {Bratek}, Sebastian {Kubis},
  and Tomasz {Kedziorek}.
\newblock {Oscillating Strange Quark Matter Objects Excited in Stellar
  Systems}.
\newblock {\em \apj}, 897(2):168, July 2020.

\bibitem{2021PhRvD.103j3003S}
Ting-Ting {Sun}, Zi-Yue {Zheng}, Huan {Chen}, G.~Fiorella {Burgio}, and
  Hans-Josef {Schulze}.
\newblock {Equation of state and radial oscillations of neutron stars}.
\newblock {\em \prd}, 103(10):103003, May 2021.

\bibitem{2023PhRvD.108f4007C}
Kenneth {Chen} and Lap-Ming {Lin}.
\newblock {Fully general relativistic simulations of rapidly rotating quark
  stars: Oscillation modes and universal relations}.
\newblock {\em \prd}, 108(6):064007, September 2023.

\bibitem{2010PhLB..691...99H}
Wen-Biao {Han}, Remo {Ruffini}, and She-Sheng {Xue}.
\newblock {Electron-positron pair oscillation in spatially inhomogeneous
  electric fields and radiation}.
\newblock {\em Physics Letters B}, 691(2):99--104, July 2010.

\bibitem{2012PhRvD..86h4004H}
W.-B. {Han}, R.~{Ruffini}, and S.-S. {Xue}.
\newblock {Electron and positron pair production of compact stars}.
\newblock {\em \prd}, 86(8):084004, October 2012.

\bibitem{1987PhRvD..36..114G}
G.~{Gatoff}, A.~K. {Kerman}, and T.~{Matsui}.
\newblock {Flux-tube model for ultrarelativistic heavy-ion collisions:
  Electrohydrodynamics of a quark-gluon plasma}.
\newblock {\em \prd}, 36:114--129, July 1987.

\bibitem{2023PhRvD.108a3002P}
Mikalai {Prakapenia} and Gregory {Vereshchagin}.
\newblock {Pauli blocking effects on pair creation in strong electric field}.
\newblock {\em \prd}, 108(1):013002, July 2023.

\bibitem{1954PhRv...94..511B}
P.~L. {Bhatnagar}, E.~P. {Gross}, and M.~{Krook}.
\newblock {A Model for Collision Processes in Gases. I. Small Amplitude
  Processes in Charged and Neutral One-Component Systems}.
\newblock {\em Physical Review}, 94:511--525, May 1954.

\bibitem{2017rkt..book.....V}
Gregory~V. {Vereshchagin} and Alexey~G. {Aksenov}.
\newblock {\em {Relativistic Kinetic Theory}}.
\newblock Cambridge University Press, 2017.

\bibitem{2003MNRAS.343L..69A}
A.~G. {Aksenov}, M.~{Milgrom}, and V.~V. {Usov}.
\newblock {Radiation from hot, bare, strange stars}.
\newblock {\em \mnras}, 343:L69--L72, August 2003.

\bibitem{2000ApJ...545L.127Z}
Bing {Zhang}, R.~X. {Xu}, and G.~J. {Qiao}.
\newblock {Nature and Nurture: a Model for Soft Gamma-Ray Repeaters}.
\newblock {\em \apjl}, 545(2):L127--L130, December 2000.

\bibitem{2013PhLA..377..206B}
A.~{Benedetti}, R.~{Ruffini}, and G.~V. {Vereshchagin}.
\newblock {Phase space evolution of pairs created in strong electric fields}.
\newblock {\em Physics Letters A}, 377:206--215, January 2013.

\bibitem{2007PhRvL..99l5003A}
A.~G. {Aksenov}, R.~{Ruffini}, and G.~V. {Vereshchagin}.
\newblock {Thermalization of Nonequilibrium Electron-Positron-Photon Plasmas}.
\newblock {\em \prl}, 99(12):125003--+, September 2007.

\bibitem{2009PhRvD..79d3008A}
A.~G. {Aksenov}, R.~{Ruffini}, and G.~V. {Vereshchagin}.
\newblock {Thermalization of the mildly relativistic plasma}.
\newblock {\em \prd}, 79(4):043008, February 2009.

\bibitem{diclemente2024strangequarkmatterdark}
Francesco~Di Clemente, Marco Casolino, Alessandro Drago, Massimiliano Lattanzi,
  and Claudia Ratti.
\newblock Strange quark matter as dark matter: 40 years later, a reappraisal,
  2024.

\bibitem{2005MNRAS.357.1377C}
E.~{Churazov}, R.~{Sunyaev}, S.~{Sazonov}, M.~{Revnivtsev}, and
  D.~{Varshalovich}.
\newblock {Positron annihilation spectrum from the Galactic Centre region
  observed by SPI/INTEGRAL}.
\newblock {\em \mnras}, 357(4):1377--1386, March 2005.

\bibitem{2021Innov...200152G}
Jinjun {Geng}, Bing {Li}, and Yongfeng {Huang}.
\newblock {Repeating fast radio bursts from collapses of the crust of a strange
  star}.
\newblock {\em The Innovation}, 2:100152, November 2021.

\bibitem{1991NuPhS..24...40C}
T.~{Chmaj}, P.~{Haensel}, and W.~{S{\l}omi{\'n}ski}.
\newblock {Photon emissivity of strange matter}.
\newblock {\em Nuclear Physics B Proceedings Supplements}, 24:40--44, December
  1991.

\bibitem{2003ApJ...596..451C}
K.~S. {Cheng} and T.~{Harko}.
\newblock {Surface Photon Emissivity of Bare Strange Stars}.
\newblock {\em \apj}, 596(1):451--463, October 2003.

\bibitem{2010PhLB..690..250Z}
B.~G. {Zakharov}.
\newblock {Effect of electric field of the electrosphere on photon emission
  from quark stars}.
\newblock {\em Physics Letters B}, 690(3):250--254, June 2010.

\bibitem{2010PhR...487....1R}
R.~{Ruffini}, G.~{Vereshchagin}, and S.-S. {Xue}.
\newblock {Electron-positron pairs in physics and astrophysics: From heavy
  nuclei to black holes}.
\newblock {\em \physrep}, 487:1--140, February 2010.

\bibitem{1982ApJ...252..337D}
J.~K. {Daugherty} and A.~K. {Harding}.
\newblock {Electromagnetic cascades in pulsars.}
\newblock {\em \apj}, 252:337--347, January 1982.

\end{thebibliography}
\bibliographystyle{unsrt}

\end{document}